\documentclass[a4paper,11pt]{article}
\usepackage{pos}
\usepackage{adjustbox}
\usepackage{graphicx}
\usepackage{float}
\usepackage{subcaption}
\usepackage{wrapfig}

\DeclareMathOperator{\SU}{\mathrm{SU}}
\DeclareMathOperator{\Z}{\mathbb{Z}}

\title{Effective String Theory of three-dimensional $\SU(N)$ gauge theories beyond the Nambu--Got{\=o} approximation}

\ShortTitle{EST of three-dimensional $\SU(N)$ gauge theories beyond NG approximation}

\author[a]{Michele Caselle}
\author[b]{Nicodemo Magnoli}
\author[a]{Alessandro Nada}
\author[a]{Marco Panero}
\author*[a]{Dario Panfalone}
\author[a]{Lorenzo Verzichelli}

\affiliation[a]{Department of Physics, University of Turin and INFN, Turin,\\
  Via Pietro Giuria 1, I-10125 Turin, Italy}

\affiliation[b]{Department of Physics, University of Genoa and INFN, Genoa,\\
Via Dodecaneso 33, I-16146, Genoa, Italy}

\emailAdd{dario.panfalone@unito.it}

\abstract{
We study the effective bosonic string that describes confining flux tubes in three-dimensional $\SU(N)$ Yang--Mills theories. Although the low-energy properties are universal and well described by the Nambu--Got{\=o} action, the subtle dependence on the gauge group is embedded in a series of corrections, which remain undetermined, appearing in the expansion around the limit of an infinitely long string. We extract the first two of these corrections from a set of high-precision Monte Carlo simulations of Polyakov loop correlators at finite temperatures close to the deconfinement transition. We present and compare the results of new lattice simulations for theories with $N=3$ and $N=6$ color charges, along with an improved estimate for the $N=2$ case, discussing the approach to the large-$N$ limit. We show that our results are compatible with analytical bounds derived from the S-matrix bootstrap approach. Additionally, we present a new test of the Svetitsky--Yaffe conjecture for the $\SU(3)$ theory in three dimensions, showing that our results for the correlator of Polyakov loops perfectly agree with the predictions obtained using a conformal perturbation approach to the two-dimensional three-state Potts model.
}

\FullConference{%
  The 41st International Symposium on Lattice Field Theory (LATTICE2024)\\
  28 July - 3 August 2024\\
  Liverpool, UK
}

\begin{document}
\maketitle

\section{Introduction and motivation}
The non-perturbative confining properties of Yang--Mills theories can be precisely described and modelled in the Effective String Theory (EST) approach, where the flux tubes connecting a quark--antiquark pair is represented as a thin, vibrating bosonic string~\cite{Luscher:1980ac, Luscher:1980fr}. 

This string-like representation of the confining flux tube between color charges at large distances results in the emergence of a linearly confining potential, and the Polyakov loop correlator can be written in terms of an effective string partition function

\begin{equation}
            G(R) \sim \int \mathrm{D} X e^{-S_{EST}(X)} \equiv Z_{EST}.
\end{equation}
The simplest Poincar\'e-invariant EST is the Nambu--Got{\=o} (NG) string model, where the string action $S_{\mathrm{NG}}$ is defined as:
\begin{align}
 S_{\mathrm{NG}}=\sigma_0 \int_{\Sigma} \mathrm{d}^2\xi \sqrt{g},
\end{align}
with $g \equiv \mathrm{det} g_{\alpha\beta}$ and $ g_{\alpha\beta} = \partial_\alpha X_\mu \partial_\beta X^\nu$ being the metric induced on the reference world-sheet surface $\Sigma$; $\xi \equiv (\xi^0,\xi^1)$ are the world-sheet coordinates and $\sigma_0$ is the zero-temperature string tension.

Fixing its reparametrization invariance of coordinates, the string action can be expressed as a low-energy expansion in the number of derivatives of the transverse degrees of freedom of the string which, by an appropriate redefinition of the fields, can be rephrased as a large distance expansion.
In this framework, the partition function can be written explicitly, and the Polyakov loop correlator, the object of interest in this work, can be expressed in terms of a series of modified Bessel functions of the second kind.

At large distances, the correlator $G(R)$ is dominated by the ground state ($n=0$) and can be approximated as
\begin{equation}
\label{eq:correlator_NG}
G(R) \sim  K_{0} (E_0 R),
\end{equation}
with the lowest energy state given by
\begin{equation}
\label{eq:ground_NG}
E_0 = \sigma_0 L_t \sqrt{1 - \frac{\pi}{3 \sigma_0 L_t^2}}
\end{equation}
where $L_t$ is the extent of the Euclidean-time direction. It is important to notice that $E_0$ can be interpreted as the inverse of the correlation length $\xi_l$.

In this picture, it is natural to define the NG deconfinement critical temperature $T_{c,\mathrm{NG}}$ ~\cite{Pisarski:1982cn, Olesen:1985ej} as the temperature at which $E_0$ is vanishing:

\begin{equation}
    \frac{T_{c,\mathrm{NG}}}{\sqrt{\sigma_0}} = \sqrt{\frac{3}{\pi}}.
\end{equation}
From this, it is possible to predict the critical exponent $\nu = 1/2$. 

However, this description of the deconfinement transition is likely incorrect, as the critical exponent should instead correspond to that of the symmetry-breaking phase transition in the $(D-1)$-dimensional spin model, with the symmetry group being the center of the original gauge group. Moreover, this prediction is close to the correct one but quantitatively wrong, as we will analyze later. 

These observations suggest that the Nambu--Got{\=o} action should be considered merely as a leading-order approximation of the true Effective String Theory that describes the infrared dynamics of confining gauge theories. Recent advancements in flow-based architectures \cite{Caselle:2024ent} have demonstrated their effectiveness in numerically investigating these theories, enabling the study of terms beyond the Nambu--Got{\=o} action. The determination of the terms beyond this approximation is still an open problem in this context and a quantitative analysis of these terms would provide valuable insights into the characteristics of different confining gauge theories.

This contribution concerns our recent work~\cite{Caselle_2024} where, following the same strategy used in ref.~\cite{Caristo:2021tbk}  for the $\SU(2)$ gauge theory, we studied the fine details of the large-distance behavior of the Polyakov loop correlator of the $\SU(3)$ and $\SU(6)$ gauge theories in $D=2+1$ spacetime dimensions.

\section{Detection of beyond NG corrections from lattice simulations}
In recent studies it has been found a crucial feature characterizing the EST, known as ``low-energy universality''~\cite{Luscher:2004ib, Meyer:2006qx, Aharony:2009gg, Aharony:2011gb, Gliozzi:2011hj, Gliozzi:2012cx, Dubovsky:2012sh, Aharony:2013ipa}, which originates directly from the symmetry constraints imposed by Poincaré invariance in the target space. According to this property, in the high-temperature regime, the first correction with respect to the NG action appears at order $1/L_t^{7}$ in the expansion of $E_0$ around the limit of an infinitely long string. This explains the remarkable success of the Nambu--Got{\=o} model in describing the infrared behavior of confining gauge theories, despite its simplicity. At the same time, it highlights why identifying corrections beyond the Nambu--Got{\=o} (BNG) approximation is so challenging. 

This same result can also be obtained in an independent way, using a bootstrap type of analysis: this was done in refs.~\cite{EliasMiro:2019kyf, EliasMiro:2021nul} in the framework of the S-matrix approach. Moreover, it was discovered that the first two correction terms beyond the Nambu--Got{\=o} approximation (the $1/L_t^7$ and $1/L_t^9$ terms) are controlled by the same parameter, while the next independent parameter only appears at the $1/L_t^{11}$ order.

Therefore, the expression that we will compare with the results of our simulations for the $N_t=L_t/a$ dependence of the ground state $E_0$ is the following:
\begin{align}
\label{eq:BNG_correction}
\begin{aligned}
   aE_0(N_t) = & N_t\sigma_0 a^2 \sqrt{1-\frac{\pi}{3N_t^2 \; \sigma_0 a^2}} + \frac{k_4}{(\sigma_0 a^2)^3 N_t^7} + \frac{2 \pi k_4}{3(\sigma_0 a^2)^4 N_t^9} + \\
   & \frac{5 \pi^2 k_4}{16 (\sigma_0 a^2)^5 N_t^{11}} + \frac{k_5}{(\sigma_0 a^2)^5 N_t^{11}} + \dots
\end{aligned}
\end{align} 

Since there may also be higher-order corrections, it is natural to truncate for consistency the Nambu--Got{\=o} prediction to the corresponding order of the correction. Moreover, the values of the parameters $k_4$ and $k_5$ are strongly affected by both the order of the last correction that is included and the order at which we truncate the Nambu--Got{\=o} prediction. For this reason, to obtain a reliable estimate of these parameters we carried out a careful analysis of this systematic.

\section{Numerical results}
Following the same approach of the study of the $\SU(2)$ gauge theory in $(2+1)$ dimensions ~\cite{Caristo:2021tbk}, we estimated the EST correction beyond the Nambu--Got{\=o} approximation in the $\SU(3)$ and $\SU(6)$ lattice gauge theories, by analyzing the behavior of the ground-state energy $E_0$ as the deconfinement transition temperature is approached from below. In our lattice simulations we computed the values of $G(R)$ for $R \le L_s/2$ and, at a fixed value of $\beta$, the temperature is changed by varying $N_t$. We run numerical simulations at different values of $\beta$ and we determined the ground state energy $E_0$ as the inverse of the longest correlation length in the system, denoted as $\xi_l$. 

\subsection{$\SU(3)$ Yang--Mills theory}
In the $\SU(3)$ case we evaluated the correlation length in two independent ways. First, by performing a long distance fit of the Polyakov correlator assuming the EST functional form 
\begin{equation}
    G(R) = k_l \, \left[K_0\left(\frac{R}{\xi_l}\right) + K_0\left(\frac{L_s - R}{\xi_l}\right)\right].
\end{equation}

The second, following the procedure of ref.~\cite{Caristo:2021tbk} for the $\SU(2)$ gauge theory, by starting from the Svetitsky--Yaffe conjecture~\cite{Svetitsky:1982gs} which maps the $\SU(3)$ Yang--Mills theory in three dimensions (which, like the $\SU(2)$ theory, also has a continuous thermal deconfinement phase transition) to the three-state Potts model in two dimensions. We fitted our data for the Polyakov loop correlator using the form of the spin-spin correlator predicted from the Potts model using conformal perturbation theory ~\cite{Caselle:2005sf,GUIDA1996361}.

The values of the correlation length $\xi_l$ obtained from the short-range fits are in good agreement (within less than three standard deviations) with those obtained by fitting the large distance data. This agreement confirms the robustness of our determination of the ground-state energy, gives confidence on the procedure to determine BNG corrections, and can be seen as a further confirmation of Svetitsky--Yaffe mapping.

\begin{figure}
\centering
    \begin{subfigure}{0.56\textwidth}
        \includegraphics[width=\textwidth]{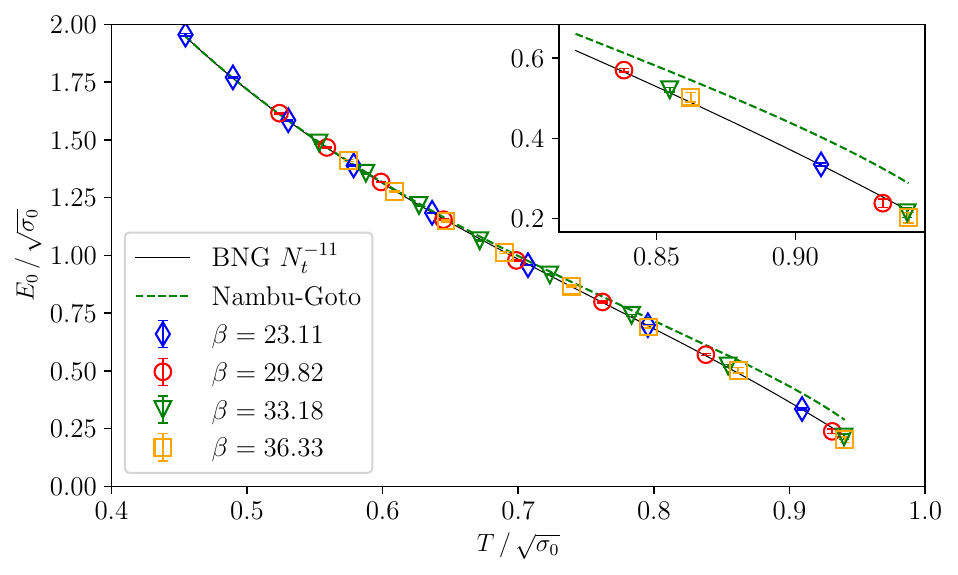}
        \caption{}
        \label{fig:combined_fit_SU3_Nt11}
    \end{subfigure}
    \begin{subfigure}{0.42\textwidth}
        \includegraphics[width=\textwidth]{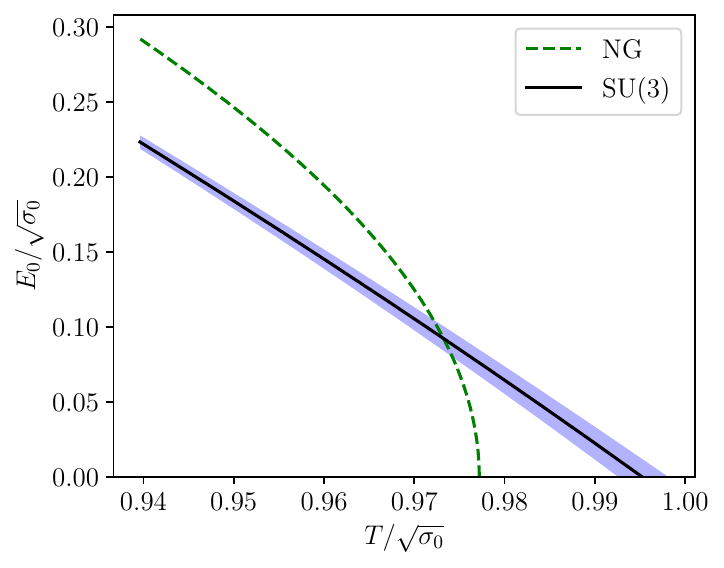}
        \caption{}
        \label{fig:NG_vs_SU3}
    \end{subfigure}
\caption{\textbf{(a)} Combined best fits of the $\SU(3)$ ground state energy $E_0$ for different values of $\beta$. The data are expressed in units of $\sqrt{\sigma_0}$ on both axes. In the zoomed inset we highlight the closest points to the critical temperature where the discrepancy between our data and the NG prediction becomes most pronounced.
\textbf{(b)} Detail of the dependence of the ground state energy $E_0$ on the temperature very close to $T_c$ for the $\SU(3)$ gauge theory, using our best fit parameters for the extrapolation (solid line with statistical confidence band).  
}
\end{figure}

To estimate the corrections we performed the best fits of the ground state energy according to the model \eqref{eq:BNG_correction}, adding the correction order by order truncating the Nambu--Got{\=o} series expansion at the term corresponding to the highest-order correction. Given the absence of any clear trend when the lattice spacing is made finer, we performed a combined fit across all values of $\beta$ fixing the same value of $k_4$ and $k_5$ for all data sets, leaving the values of $\sigma_0 a^2$ as the remaining four independent free parameters. This is a significant consistency check, indicating that the scale dependence of these coefficients has already been taken into account by the $1/\sigma_0^3$ and $1/\sigma_0^4$ normalizations in the respective terms of eq.~\eqref{eq:BNG_correction}. Furthermore, the values we obtained for the string tension at zero temperature perfectly agree with the determinations from ref.~\cite{Teper:1998te}. For instance, the combined fit up to the order $N_t^{-11}$ is illustrated in fig.~\ref{fig:combined_fit_SU3_Nt11}, where the data points systematically lie below the NG curve, hence the negative value of $k_4$ in the fit.

Moreover, it is possible to extrapolate the value of the ground state energy $E_0$ to higher temperatures, using the model truncated at order $N_t^{-11}$ and the values of $k_4$ and $k_5$ obtained from the best fit of the data. Following this procedure, we find that the model line crosses the NG prediction at a temperature around $T = 0.973(5) \sqrt{\sigma_0} \lesssim T_{c,\mathrm{NG}}$, as shown in fig.~\ref{fig:NG_vs_SU3}. The model then reaches the horizontal axis (which we expect in correspondence of the second order phase transition) at $T_{E_0 = 0} = 0.995(5) \sqrt{\sigma_0}$, which is in good agreement with the critical value $T_c = 0.9890(31)\sqrt{\sigma_0}$ reported in ref.~\cite{Liddle:2008kk}.

% \begin{figure}
% \centering
% \includegraphics[width=0.6\textwidth]{extrp_ground3.pdf}
% \caption{Detail of the dependence of the ground state energy $E_0$ on the temperature very close to $T_c$ for the $\SU(3)$ gauge theory, using our best fit parameters for the extrapolation (solid line with statistical confidence band). It is clear that the line crosses the NG one before the critical temperature and reaches the $E_0 = 0$ axis at a temperature which is compatible with the critical temperature of the theory.}
% \label{fig:NG_vs_SU3}
% \end{figure}

However, the value of the coefficients shows significant fluctuations from different combined fits, depending on the order of corrections included and the order at which the Nambu--Got{\=o} prediction is truncated. These variations suggest that the systematic error due to the truncation of the series is the primary source of uncertainty in our results for $k_4$ and $k_5$. In order to evaluate this systematic uncertainty, we repeated the fit for each order both with and without truncating the NG series at the order corresponding to the finest correction beyond Nambu--Got{\=o}. Comparing these results we observed a better agreement between these two prescriptions when higher-order corrections are included. As our final central value, we choose the result obtained truncating consistently both the NG series and the corrections at the $1/N_t^{11}$ order. The systematic error is chosen in order to include the $k_4$ values obtained truncating the BNG corrections at the $1/N_t^9$ and $1/N_t^{11}$ orders under both prescriptions.

% \begin{figure}
% \centering
% \includegraphics[width=0.6\textwidth]{k4_systematic_SU3.pdf}
% \caption{Values of the $k_4$ for the $\SU(3)$ theory obtained through various combined fits, considering corrections beyond the Nambu--Got{\=o} string up to orders $1/N_t^7$, $1/N_t^9$ and $1/N_t^{11}$, both truncating the NG baseline to the order consistent with the finest correction (blue circles) and keeping all the NG orders (red squares). We also show our final estimate, with the systematic confidence band (green band).}
% \label{fig:k4_syst_SU3}
% \end{figure}

We quote as a final result for the $\SU(3)$ theory:
\begin{equation}
    k_4 = -0.102 (11) [50], \;\;\; k_5 = 0.45 (8) [25],
\end{equation}
where the value in parentheses denotes the statistical error, while the value in square brackets indicates the systematic error.

\subsection{$\SU(6)$ Yang--Mills theory}
\begin{figure}
 \centering 
 \includegraphics[width=1\textwidth]{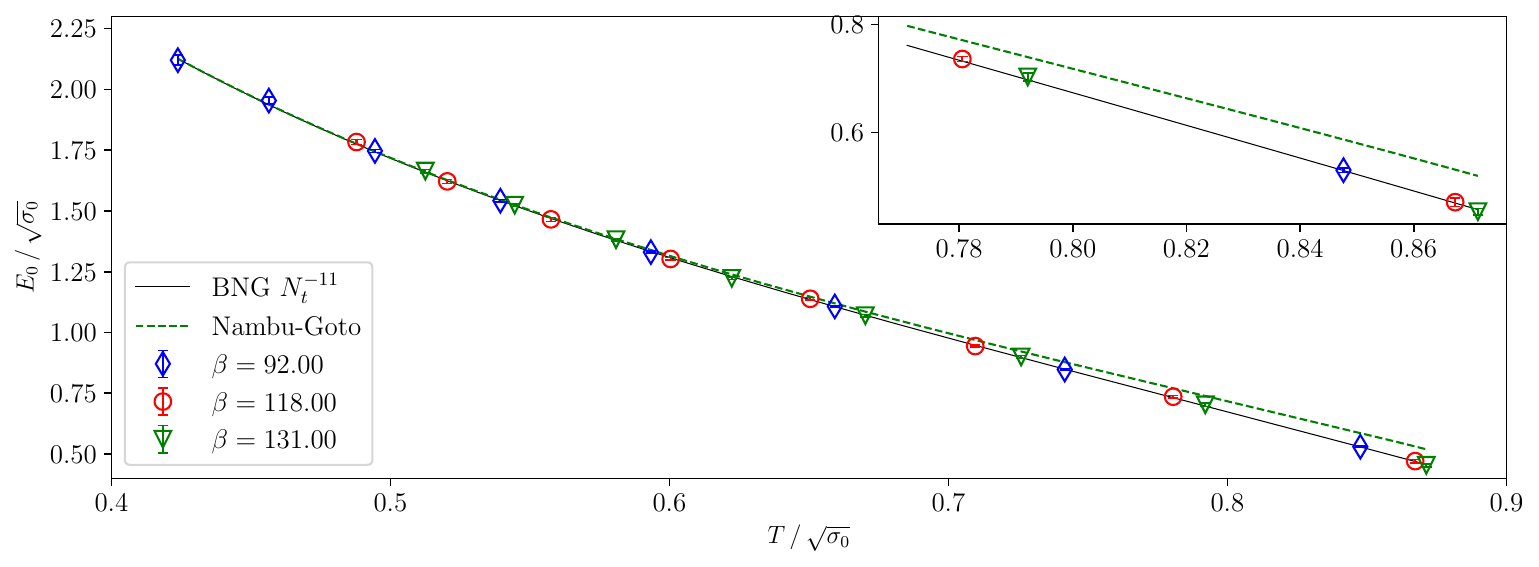}
 \caption{Combined best fits of the ground-state energy $E_0$ in the $\SU(6)$ theory, for different values of $\beta$, according to eq.~\eqref{eq:BNG_correction} including all terms up to $1/N_t^{11}$.}
 \label{fig:combined_fit_SU6_Nt11}
\end{figure}

The study of the results for the SU(6) Yang--Mills theory follows closely the procedure applied in the SU(3) case, both regarding the numerical simulations and the analysis. In this case, we determined the ground state energy solely from the long-distance fit, as the Svetitsky--Yaffe conjecture is not applicable to the SU(6) gauge theory in three dimensions, given that the phase transition is of first order~\cite{Liddle:2008kk}. The combined best fit of the ground state energy up to the order $N_t^{-11}$ is plotted in fig.~\ref{fig:combined_fit_SU6_Nt11}. Also in this case we report the final results of the first two BNG corrections both with statistical and systematic uncertainty:
\begin{equation}
    k_4 = -0.173(30)[79], \;\;\; k_5 = 0.98(23)[15].
\end{equation}

\subsection{Summary of SU(N) BNG corrections and comparison with bootstrap constraints}
\begin{figure}
    \centering
    \includegraphics[width=\textwidth]{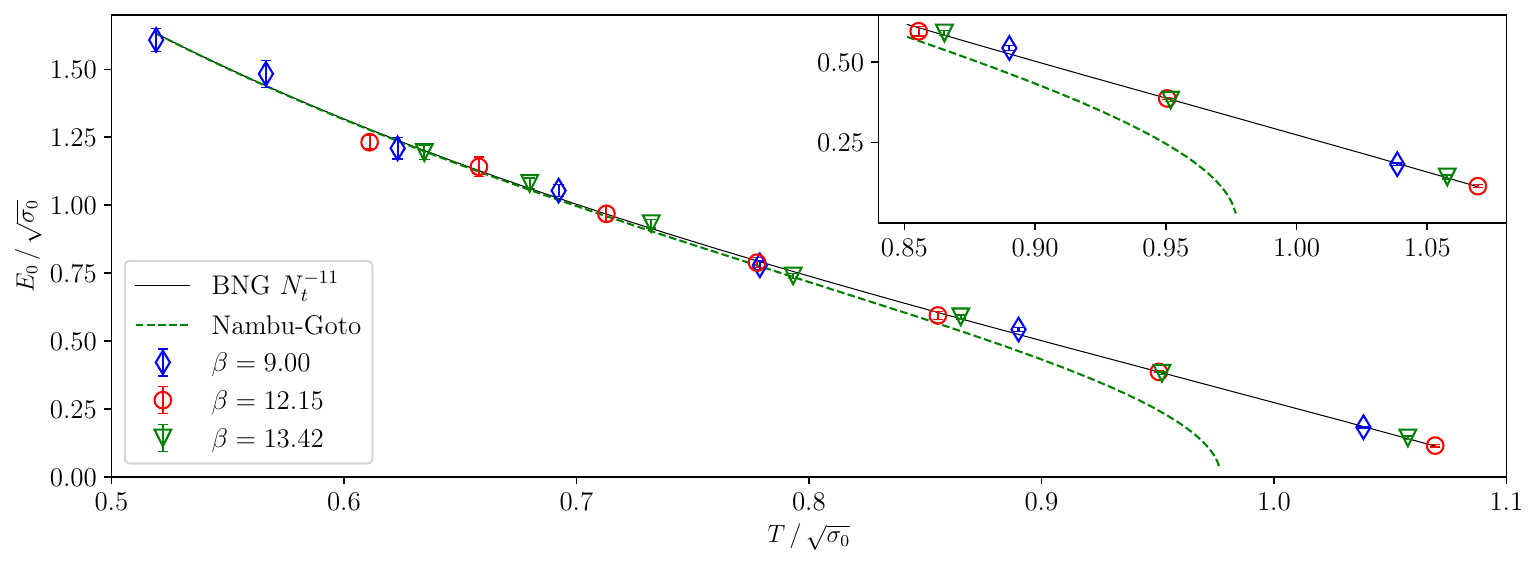}
    \caption{Combined best fits of the ground-state energy $E_0$ in the $\SU(2)$, for different values of $\beta$, according to eq.~\eqref{eq:BNG_correction} including all terms up to $1/N_t^{11}$.}
    \label{fig:combined_fit_SU2}
\end{figure}

In order to compare our results with the values previously obtained for SU(2) Yang--Mills theory we obtained, from a reanalysis of data published in ref.~\cite{Caristo:2021tbk}, an improved estimate of $k_4$ and a novel estimate of $k_5$ performing the combined fit for all lattice spacings, shown in fig.~\ref{fig:combined_fit_SU2}. 

In this case, our final results are:
\begin{equation}
    k_4 = 0.0386(95) [121], \;\;\; k_5 = -0.123(52).
\end{equation}
% \begin{wrapfigure}{r}{0.5\textwidth}  
%     \centering
%     \includegraphics[width=0.5\textwidth]{k4_vs_Tc.pdf}
%     \caption{$k_4$ values for different confining gauge theories in three spacetime dimensions.}
%     \label{fig:k4_vs_Tc}
% \end{wrapfigure}
\begin{figure}
\centering
    \begin{subfigure}{0.46\textwidth}
        \includegraphics[width=\textwidth]{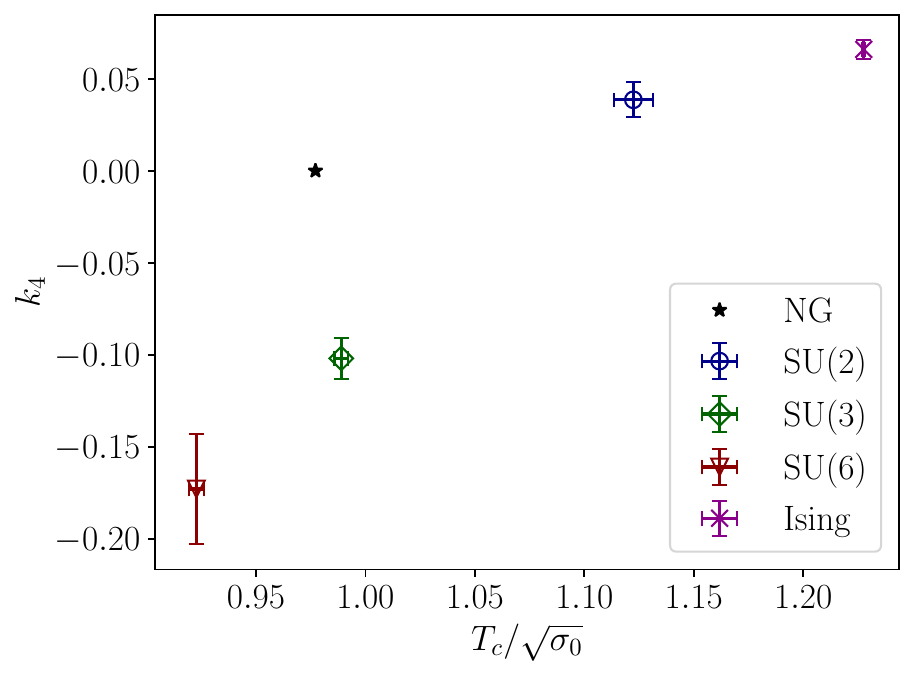}
        \caption{}
        \label{fig:k4_vs_Tc}
    \end{subfigure}
    \begin{subfigure}{0.44\textwidth}
        \includegraphics[width=\textwidth]{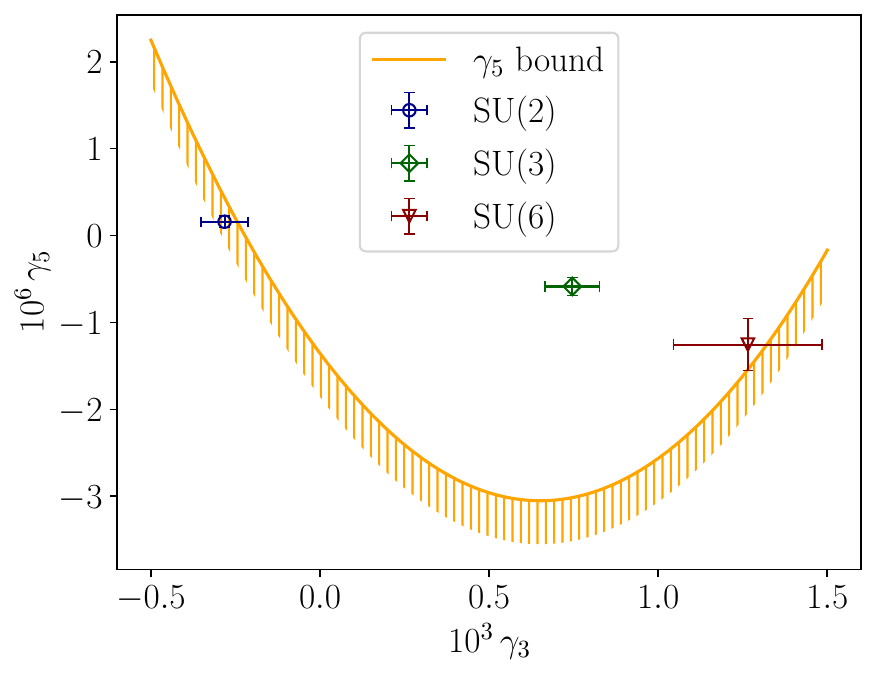}
        \caption{}
        \label{fig:gamma_bound}
    \end{subfigure}
\caption{\textbf{(a)} $k_4$ values for different confining gauge theories in three spacetime dimensions.
\textbf{(b)} Values of $\gamma_3$ and $\gamma_5$ for the $\SU(N)$ theories considered in this contribution. The solid line represents the lower bound on $\gamma_5$ obtained from the bootstrap analysis, with the shaded region indicating the forbidden area.  
}
\end{figure}

It is now insightful to compare the values of the $k_4$ coefficient across $\SU(N)$ Yang--Mills theories and with the result obtained for the $\Z_2$ gauge theory in ref.~\cite{Baffigo:2023rin}: these estimates are shown in fig.~\ref{fig:k4_vs_Tc}. Firstly, we note that the $k_4$ coefficient appears to decrease as the ``size'' of the gauge group increases. In particular, $k_4$ is positive for both the $\Z_2$ and $\SU(2)$ gauge theories, but turns negative for the $\SU(3)$ and $\SU(6)$ theories. Even though the numerical values are relatively close, the statistical and systematic uncertainties that we estimated suggest that for the $\SU(6)$ theory $k_4$ is larger in magnitude (and thus more negative).

As the last step in the analysis of the corrections of the Nambu--Got{\=o} contribution to $E_0$, we compare our results for the $k_4$ and $k_5$ parameters with the bounds found from the S-matrix bootstrap. To this purpose it is convenient, to express $k_{4,5}$ in terms of $\gamma_{3,5}$ (plotted in fig.~\ref{fig:gamma_bound}), using the same notation of ref.~~\cite{EliasMiro:2019kyf}

\begin{minipage}[t]{0.45\textwidth}  % Minipage per l'equazione a sinistra
    \vspace{-0.7cm}
    \begin{align*}
    \gamma_3 = - \frac{225}{32 \pi^6} k_4, \;\;\;\;\; \gamma_5 = - \frac{3969}{32768 \pi^{10}} k_5:
    \end{align*}
\end{minipage}%
\hfill  % Separazione tra la minipage di sinistra e quella di destra
\begin{minipage}[t]{0.5\textwidth}  % Minipage per la tabella a destra
    \centering  % Qui va bene il centering
    \begin{tabular}{|c|c|c|}
    \hline
     & $\gamma_3 \times 10^3$ & $\gamma_5 \times 10^6$ \\
    \hline
    $\SU(2)$ & $-0.282 (70) [89] $ & $ 0.159 (66) $ \\
    \hline
    $\SU(3)$ & $ 0.746 (80) [365]$ & $-0.58 (11) [32]$ \\
    \hline
    $\SU(6)$ & $ 1.26 (22) [58] $ & $-1.25 (30) [19]$ \\
    \hline
    \end{tabular}
\end{minipage}
\newline
% \begin{wrapfigure}{r}{0.5\textwidth}  
%     \centering
%     \includegraphics[width=0.5\textwidth]{gamma_bound.pdf}
%     \caption{Values of $\gamma_3$ and $\gamma_5$ for the $\SU(N)$ theories considered in this contribution. The solid line represents the lower bound on $\gamma_5$ obtained from the bootstrap analysis, with the shaded region indicating the forbidden area.}
%     \label{fig:gamma_bound}
% \end{wrapfigure}

Firstly, all values of $\gamma_3$ are well inside the bound $\gamma_3 > - \frac{1}{768} \simeq -0.0013$. Moreover, the bound on $\gamma_5$ depends on the one on $\gamma_3$ and it is denoted by the solid yellow line shown in fig.~\ref{fig:gamma_bound}. The $\gamma_5$ values for $\SU(3)$ and $\SU(6)$ from our analysis are in the allowed region, while the result for $\SU(2)$ is slightly outside this region. However, when considering both statistical and systematic uncertainties, the $\SU(2)$ result may still be compatible with the bound.

The values we obtained for the $\SU(3)$ and $\SU(6)$ theories are slightly higher than those predicted in refs.~\cite{Guerrieri:2024ckc,Dubovsky:2014fma}. In particular our $\SU(6)$ value is on the edge of the most constraining bound in their \emph{branon matryoshka}. However, this bound was derived using the condition $\gamma_3 < 10^{-3}$, which our result satisfies within 1.2 statistical errors.

The $\gamma_3$ values we obtained for the $\SU(3)$ and $\SU(6)$ theories are quite similar, indicating a weak dependence on $N$ for $N \geq 3$. This trend would be in line with what is typically observed for other observables in $\SU(N)$ Yang--Mills theories, both in $2+1$ and in $3+1$ dimensions~\cite{Lucini:2012gg, Panero:2012qx}. Assuming a dependence of $\gamma_3$ on the number of color charges of the form
\begin{align}
\label{gamma3_vs_N}
\gamma_3^{(N = \infty)} + \frac{c}{N^2},
\end{align}
we performed a best fit using our numerical data from the $\SU(3)$, $\SU(6)$, and $\SU(2)$ theories. The result is $\gamma_3^{(N = \infty)}  =1.54(13)\times 10^{-3}$, which is within one standard deviation from our result for the $\SU(6)$ Yang--Mills theory.

% \begin{figure}
%     \centering
%     \begin{subfigure}{0.49\textwidth}
%         \includegraphics[width=\textwidth]{NG_VS_BNG_SU3.pdf}
%         \caption{}
%         \label{fig:combined_fit_SU2}
%     \end{subfigure}
%     \hfill
%     \begin{subfigure}{0.49\textwidth}
%         \includegraphics[width=\textwidth]{NG_VS_BNG_SU6.pdf}
%         \caption{}
%         \label{fig:combined_fit_SU3}
%     \end{subfigure}
%     \hfill
%     \begin{subfigure}{0.49\textwidth}
%         \includegraphics[width=\textwidth]{NG_VS_BNG_SU2.pdf}
%         \caption{}
%         \label{fig:combined_fit_SU6}
%     \end{subfigure}
%     \caption{Combined best fits of the ground-state energy $E_0$ in the $\SU(2)$ theory \textbf{(a)}, $\SU(3)$ \textbf{(b)} and $\SU(6)$ \textbf{(c)}, for different values of $\beta$, according to eq.~\eqref{eq:BNG_correction} including all terms up to $1/N_t^{11}$. The data are expressed in units of $\sqrt{\sigma_0}$ on both axes. In the zoomed insets we highlight the closest points to the critical temperature where the discrepancy between our data and the NG prediction becomes most pronounced.}
%     \label{fig:combined_fit_SU6_Nt11}
% \end{figure}
\section{Conclusion}
In this contribution we presented a systematic study of the effective string corrections beyond the Nambu--Got{\=o} action, for $\SU(N)$ Yang--Mills theories in $2+1$ spacetime dimensions. One of the main goals of our analysis consisted in investigating the fine details characterizing the confining dynamics in theories based on different gauge groups.

Our analysis was based on a new set of high-precision results of Monte Carlo simulations of the two-point Polyakov loop correlation function, determined non-perturbatively in the lattice regularization. Following the approach used in ref.~\cite{Caristo:2021tbk} for the $\SU(2)$ gauge theory, we extended the investigation to the cases of the theories with $N=3$ and $6$ color charges in the proximity of their deconfinement phase transition at finite temperature. From these simulations, we extracted the ground-state energy of the effective string and estimated deviations from the Nambu--Got{\=o} approximation. These deviations were parameterized in terms of the $k_4$ and $k_5$ coefficients, which appear in the expansion of the correlator in the long-string limit.

Moreover, we provided an improved estimate of the $k_4$ coefficient and a new determination of the $k_5$ coefficient for the $\SU(2)$ Yang--Mills theory, using the data reported in ref.~\cite{Caristo:2021tbk}, with an improved evaluation of systematic effects. Our estimated coefficients can be directly translated in terms of the $\gamma_3$ and $\gamma_5$ coefficients, and are found to be in agreement with the bounds obtained from the bootstrap analysis.

In parallel, we conducted a new high-precision test of the Svetitsky--Yaffe conjecture~\cite{Svetitsky:1982gs} by comparing the analytical solution of the short-distance spin-spin correlator from the two-dimensional three-state Potts model with our data for the $\SU(3)$ gauge theory in $2+1$ dimensions at finite temperature. The functional form predicted from the Potts model by means of conformal perturbation theory was successfully fitted to the results for the Polyakov loop correlator of the gauge theory at short distances, and the fit yields a correlation length in remarkable agreement with the one obtained from the long-range fits motivated by the EST predictions.

\subsection*{Acknowledgements}

We thank A.~Bulgarelli and E.~Cellini for helpful discussions. This work was partially supported by the Simons Foundation grant
994300 (Simons Collaboration on Confinement and QCD Strings) and by the
Prin 2022 grant 2022ZTPK4E. The simulations were run on CINECA computers. We acknowledge support from the SFT Scientific Initiative of the Italian Nuclear Physics Institute (INFN).

\bibliographystyle{JHEP}

\providecommand{\href}[2]{#2}
\end{document}